\documentclass[aps,showpacs,twocolumn]{revtex4}
\usepackage{epsfig}
\usepackage{color}
\usepackage{amsmath}
\usepackage{amssymb}
\usepackage{bbm}

\newcommand{\be}{\begin{equation}}
\newcommand{\ee}{\end{equation}}
\newcommand{\bea}{\begin{eqnarray}}
\newcommand{\beaa}{\begin{eqnarray*}}
\newcommand{\eea}{\end{eqnarray}}
\newcommand{\eeaa}{\end{eqnarray*}}

\begin{document}

\title{\bf\Large {Numerical solution of Maxwell equations for s-wave superconductors}}

\author{Naoum Karchev and Tsvetan Vetsov}

\affiliation{Department of Physics, University of Sofia, 1164 Sofia, Bulgaria}

\begin{abstract}

We report the numerical solutions of the system of equations, which describes the electrodynamics of s-wave superconductors, for time independent fields and half-plane superconductor geometry. The results are: i)the applied magnetic field increases the Ginzburg-Landau (GL) coherence length and suppresses the superconductivity, ii)the applied electric field decreases GL coherence length and supports the superconductivity, iii) if the applied magnetic field is fixed and the applied electric field increases
the London penetration depth of the magnetic field decreases. The main conclusion is that applying electric field at very low temperature  one increases the critical magnetic field. This result is experimentally testable.

\end{abstract}

\pacs{74.20.-z,74.20.Mn,71.10.-w}

\maketitle

There is at present no general understanding of the interplay between applied electric, magnetic fields and superconductivity. The London equations \cite{London35} explain the Meissner-Ochsenfeld effect but  can not predict the interplay between applied electric, magnetic fields and superconductivity (density of Cooper pairs). In paper \cite{Hirsch03} the electric field in interior of superconductors is justified using  theory of hole superconductivity \cite{Hirsch89}.

The experiments to detect an electric field in superconductors \cite{London36} are unsuccessful.
It is very difficult to experimentally test the effects of applied electric field. Right below the superconductor critical temperature the normal fluid dominates the system. The screening length of the normal fluid is about one Angstr\"{o}m. The applied field cannot penetrate into the system more than one Angstr\"{o}m from the surface. This is why the electric field cannot affect the system in the interior.  To study the effects of the applied electric field one has to do experiments at very low temperatures where there are no normal quasiparticles.

The phenomenological Ginzburg-Landau theory \cite{GL50} is a basic tool for theoretical investigation of superconductivity. A complex function $\psi(\bold x)$ is introduced as an order parameter with $|\psi(\bold x)|$ representing the local density of Cooper pairs. The free-energy density is assumed to be functional of the order parameter and the magnetic vector potential.  Using a variational method to the free-energy a system of equations, which generalized the London theory, is obtained \cite{Tinkham75,FetterWalecka}.

To derive a system of equations which includes the electric field one has to consider time dependent Ginzburg-Landau theory. We consider a relativistically covariant theory in terms of gauge four-vector electromagnetic potentials and scalar complex field-order parameter. The electrodynamics is a Lorentz covariant theory and one expects that the model under consideration will help to get deeper insight for the interplay between electric, magnetic fields and superconductivity. We want also to compare our results with the results in \cite{Hirsch04} where relativistically covariant theory of superconductivity is discussed.

The system of equations which describes the electrodynamics of s-wave superconductors reads:
\bea\label{MSc105}
& & \overrightarrow{\nabla}\times\textbf{B}\,=\,\mu\varepsilon\frac {\partial \textbf{E}}{\partial t}-2e^{*2}\rho^2\textbf{Q} \label{MSc101}\\
& & \overrightarrow{\nabla}\times\textbf{Q}\,=\,\textbf{B}\label{MSc102}\\
& & \overrightarrow{\nabla}\cdot\textbf{E}\,=\,-2e^{*2}\rho^2Q \label{MSc103}\\
& & \overrightarrow{\nabla} Q+\frac {\partial \textbf{Q}}{\partial t} \,= \,-\textbf{E}\label{MSc104}\\
& & \mu\varepsilon\frac {\partial^2 \rho}{\partial t^2}-\Delta\rho -\alpha \rho+g\rho^3- e^{*2}\rho\left [\mu\varepsilon Q^2-\textbf{Q}^2\right]=0.\nonumber\\ \label{MSc105}
\eea
where $\textbf{E}$ is the electric field, $\textbf{B}$ is the magnetic field and $\rho=|\psi|$ is the local density of Cooper pairs. The parameter $\mu$ is the magnetic permeability and $\varepsilon$ is the electric permittivity of the superconductor. We assume that they do not change their values when the system undergoes normal to superconductor transition. The parameter
\be\label{MSc12}
\alpha=\alpha_0(T_c-T),\ee where $T$ is the temperature and $T_c$ is the critical temperature, is positive when the system is superconductor. The charge of the Cooper pair is $e^{*}$

The vector $\bold{Q}$ and scalar $Q$ are supplementary fields. It is important to stress that the gauge invariant vector $\textbf{Q}$ and scalar $Q$ fields take part in the equations (\ref{MSc102}) and (\ref{MSc104}) as a magnetic vector and electric scalar potentials, while in equation (\ref{MSc101}) $(-2e^{*2}\rho^2\textbf{Q})$ is a supercurrent and in equation (\ref{MSc103}) $(-2e^{*2}\rho^2Q)$ is a density of superconducting quasi-particles . This dual contribution of the new fields is the basis of the electrodynamics of superconductors.

One can derive the system of equations (\ref{MSc101}-\ref{MSc105}) from relativistically covariant theory of superconductivity \cite{Karchev16}.

We focus on the system of equations with time-independent fields:
\bea
& & \overrightarrow{\nabla}\times\textbf{B}\,=\,-2e^{*2}\rho^2\textbf{Q} \label{MSc111}\\
& & \overrightarrow{\nabla}\times\textbf{Q}\,=\,\textbf{B}\label{MSc112}\\
& & \overrightarrow{\nabla}\cdot\textbf{E}\,=\,-2e^{*2}\rho^2Q \label{MSc113}\\
& & \overrightarrow{\nabla} Q\,=\,-\textbf{E}\label{MSc114}\\
& & \Delta\rho +\alpha \rho-g\rho^3+ e^{*2}\rho\left [\mu\varepsilon Q^2-\textbf{Q}^2\right]=0.  \label{MSc115}
\eea

It is important to say that the system of equations for static electrodynamics is not split into systems of equations for electric and magnetic fields. One can do this assuming that  $\textbf{Q}$ and $Q$ are zero in equation (\ref{MSc115}) and $\rho=\rho_0$ is a constant determined from the equation $\alpha \rho_0-g\rho_0^3=0$, which follows from the same equation (\ref{MSc115}). It is easy to obtain the equations for the electric and magnetic fields within this approximation
\bea\label{MSc1201}
& & \Delta \textbf{E}\,=\,\frac {1}{\lambda_L^2}\textbf{E} \\
& & \Delta \textbf{B}\,=\,\frac {1}{\lambda_L^2}\textbf{B}.
\eea
They imply that an electric field penetrates a distance
\be\label{MSc1202}\lambda_L=\sqrt{g/(2e^{*2}\alpha)},\ee
as a magnetic field does \cite{Hirsch04}.

This approximation is very rough and does not account for the last term in the equation (\ref{MSc115}) which is responsible for the different impact on superconductivity of applied electric and magnetic fields.

To elucidate the interplay between electric, magnetic fields and superconductivity we consider the system of equations (\ref{MSc111}-\ref{MSc115}) for fields which depend on $z$ coordinate only. Then, the system of equations for the fields $Q(z)$, $\textbf{Q}(z)=(0,Q_y(z),0)$, $\textbf{E}(z)=(0,0,E_z(z))$, $\textbf{B}(z)=(B_x(z),0,0)$ and $\rho(z)$ adopts the form
\bea
& & \frac {dB_x}{dz}\,=\,-2e^{*2}\rho^2Q_y \label{MScApp1}\\
& & \frac {dQ_y}{dz}\,=\,-B_x\label{MScApp2}\\
& & \frac{dE_z}{dz}\,=\,-2e^{*2}\rho^2Q \label{MScApp3}\\
& & \frac{dQ}{dz}\,=\,-E_z\label{MScApp4}\\
& & \Delta\rho +\alpha \rho-g\rho^3+ e^{*2}\rho\left [\mu\varepsilon Q^2-Q_y^2\right]=0.  \label{MScApp5}
\eea
After some calculations one reduces the system (\ref{MScApp1}-\ref{MScApp5}) to a system of equations for $Q,Q_y$ and $\rho$
\bea
& & \frac {d^2Q}{dz^2}\,=\,2e^{*2}\rho^2Q \label{MScApp6}\\
& & \frac {d^2Q_y}{dz^2}\,=\,2e^{*2}\rho^2Q_y \label{MScApp7}\\
& & \Delta\rho +\alpha \rho-g\rho^3+ e^{*2}\rho\left [\mu\varepsilon Q^2-Q_y^2\right]=0.  \label{MScApp8}
\eea

It is convenient to introduce dimensionless functions $f_1(\zeta),f_2(\zeta)$ and $f_3(\zeta)$ of a dimensionless distance  $\zeta=z/\xi_{GL}$, where
\be\label{MScApp9}\xi_{GL}=1/\sqrt{\alpha}\ee
is the Ginzburg-Landau coherence length:
\bea\label{MScApp9}
Q(\zeta) & = & -E_0 \xi_{GL}f_1(\zeta) \nonumber \\
Q_y(\zeta) & = &  -B_0 \xi_{GL}f_2(\zeta)  \\
\rho(\zeta) & = & \rho_0 f_3(\zeta). \nonumber  \eea
In equations (\ref{MScApp9}) $\rho_0=\sqrt{\alpha/g}$, the applied electric field is $\textbf{E}_0=(0,0,E_0)$ and the applied magnetic field is
$\textbf{B}_0=(B_0,0,0)$. The representations of the electric and magnetic fields by means of $f_1$ and $f_2$ are the following:
\bea\label{MScApp10}
E_z(\zeta) & = & E_0 \frac {df_1(\zeta)}{d\zeta} \nonumber \\
B_x(\zeta) & = & B_0 \frac {df_2(\zeta)}{d\zeta}  \eea
The system of equations (\ref{MScApp6}-\ref{MScApp8}), rewritten in terms of the new functions, reads:
\bea\label{MScApp11}
& & \frac {d^2 f_1(\zeta)}{d\zeta^2}\,=\,\frac {1}{\kappa^2}f^2_3(\zeta)f_1(\zeta) \nonumber \\
& & \frac {d^2 f_2(\zeta)}{d\zeta^2}\,=\,\frac {1}{\kappa^2}f^2_3(\zeta)f_2(\zeta) \nonumber  \\
& & \frac {d^2 f_3(\zeta)}{d\zeta^2}\,+\,f_3(\zeta)\,-\,f_3^3(\zeta)\\
& & =\,-\, f_3(\zeta)\left [\gamma_E f_1^2(\zeta)-\gamma_B f_2^2(\zeta)\right]\nonumber
\eea
In equations (\ref{MScApp11}) $\kappa$ is the Ginzburg-Landau parameter
\be\label{MScApp12}
\kappa\,=\,\frac {\lambda_L}{\xi_{GL}},\ee
which satisfies $\kappa<1/\sqrt{2}$, for type I superconductors and $\kappa>1/\sqrt{2}$ for type II ones. The parameters $\gamma_E$ and $\gamma_B$ have the representation
\be\label{MScApp13}
\gamma_E\,=\,\frac {e^{*2}\mu\varepsilon E_0^2}{\alpha^2},\hskip 1cm \gamma_B\,=\,\frac {e^{*2}B_0^2}{\alpha^2}.\ee

For semi-infinite superconductors, with a surface of superconductor orthogonal to the $z$-axis, the boundary conditions are:
\bea\label{MScApp13}
& & \frac {df_1(0)}{d\zeta}\,=\,1 \hskip 1cm f_1(\infty)\,=\,0 \nonumber \\
& & \frac {df_2(0)}{d\zeta}\,=\,1 \hskip 1cm f_2(\infty)\,=\,0  \\
& & f_3(0)\,=\,0 \hskip 1.2cm f_3(\infty)\,=\,1 .\nonumber
\eea

If neither electric nor magnetic fields are applied the equation for the dimensionless function $f_3(\zeta)\,=\,\rho(\zeta)/\rho_0$
\be\label{MScApp14}\frac {d^2 f_3(\zeta)}{d\zeta^2}\,+\,f_3(\zeta)\,-\,f_3^3(\zeta)\,=\,0\ee
is exactly solvable \cite{Tinkham75,FetterWalecka} and the solution, for $z\geq 0$ is
\be\label{MScApp15}
f_3(\zeta)\,=\,f_3(\frac {z}{\xi_{GL}})\,=\,\tanh(\frac {z}{\sqrt{2}\xi_{GL}}). \ee

\begin{figure}[!ht]
\epsfxsize=\linewidth
\epsfbox{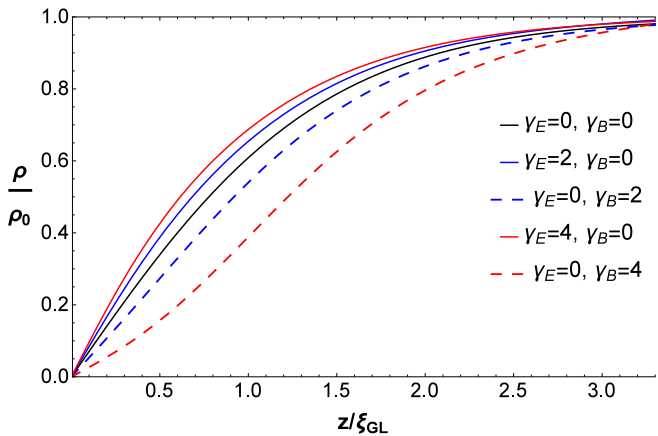} \caption{(Color online)\,\,Density of Cooper pairs $\rho/\rho_0=f_3$ as a function of $z/\zeta_{GL}$. i)dash lines-when magnetic field is applied, ii)solid lines-when the electric field is applied, iii) the line in the middle-neither electric nor magnetic fields are applied. }\label{fig1-MScII}
\end{figure}

It is more convenient to study a system of first order differential equations. To this end we introduce three new  functions
$(p_1(\zeta),\,\,\, p_2(\zeta),\,\,p_3(\zeta))$ and rewrite the system (\ref{MScApp11}) in the form
\bea\label{MScII3}
& & \frac {d p_1(\zeta)}{d\zeta}\,=\,\frac {1}{\kappa^2}f^2_3(\zeta)f_1(\zeta)\label{MScII31} \\
& & \frac {d p_2(\zeta)}{d\zeta}\,=\,\frac {1}{\kappa^2}f^2_3(\zeta)f_2(\zeta) \label{MScII32}\\
& & \frac {d p_3(\zeta)}{d\zeta}\,+\,f_3(\zeta)\,-\,f_3^3(\zeta)  \nonumber \\
& & =\,-\, f_3(\zeta)\left[\gamma_E f_1^2(\zeta)-\gamma_B f_2^2(\zeta)\right] \label{MScII33} \\
& & \frac {d f_1(\zeta)}{d\zeta}\,=\,p_1(\zeta) \label{MScII34} \\
& & \frac {d f_2(\zeta)}{d\zeta}\,=\,p_2(\zeta) \label{MScII34} \\
& & \frac {d f_3(\zeta)}{d\zeta}\,=\,p_3(\zeta) \label{MScII35}
\eea
We solve numerically the system (\ref{MScII31}-\ref{MScII35}) for $\kappa=1/3$ and different values of $\gamma_E$ and $\gamma_B$. The solutions for density of Cooper pairs $\rho/\rho_0=f_3$ as a function of $z/\zeta_{GL}$ are depicted in figure (\ref{fig1-MScII}).

The curve in the middle(black) is the solution (\ref{MScApp15}), when neither electric nor magnetic fields are applied. It is the reference solution.
The two dash curves below the reference one are solutions when magnetic field is applied ($\gamma_E=f_1=p_1=0$) and the two solid line curves above the reference one are the solutions when electric field is applied ($\gamma_B=f_2=p_2=0$).

The Ginzburg-Landau coherence length  measures the distance over which the superconducting order parameter increases up to the bulk value, measured from the surface of the superconductor ($z>0$). If we set in equation (\ref{MScApp15}) $z=\xi_{GL}\,\,(\zeta=1)$ we obtain $f_3(1)=0.6$. We can use this relation as a definition of the GL coherence length $\xi_{GL}^E$, when electric field is applied and $\xi_{GL}^B$, when magnetic field is applied. The solution $f_3^E(\zeta)$, when electric field is applied, satisfies $f_3^E(\zeta^E)=0.6$ for $\zeta^E=\xi_{GL}^E/\xi_{GL}$, and the solution $f_3^B(\zeta)$, when magnetic field is applied, satisfies $f_3^B(\zeta^B)=0.6$ for $\zeta^B=\xi_{GL}^B/\xi_{GL}$. The dash curves in figure (\ref{fig1-MScII}) show that $\zeta^B=\xi_{GL}^B/\xi_{GL}>1$ and the GL coherence length increases when applied magnetic field increases, while
$\zeta^E=\xi_{GL}^E/\xi_{GL}<1$ and the GL coherence length decreases when applied electric field increases.
In conclusion, the applied electric field decreases the GL coherence length, which means that the electric field supports the superconductivity, while the applied magnetic field increases the GL coherence length, which means that the magnetic field destroys the superconductivity.

\begin{figure}[!ht]
\epsfxsize=\linewidth
\epsfbox{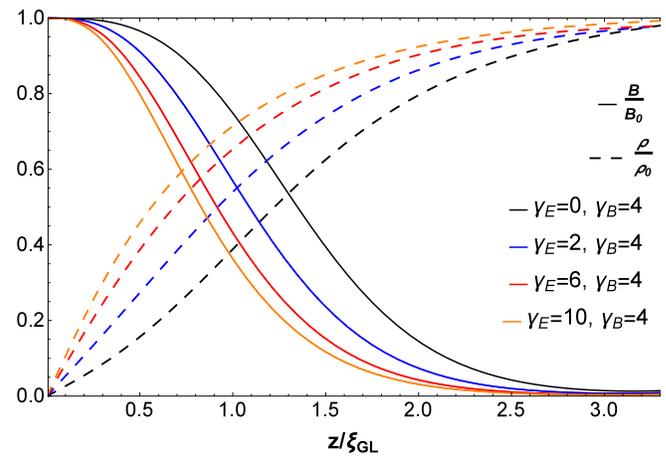} \caption{(Color online)\,\,The dash lines are the density of Cooper pairs $\rho/\rho_0=f_3$ as a function of $z/\zeta_{GL}$ for fixed value of the applied magnetic field and different values of applied electric fields. The solid lines are the magnetic field in the interior of the superconductor when different electric fields are applied.}\label{fig2-MScII}
\end{figure}

The next calculations are achieved for fixed value of the applied magnetic field $\gamma_B=4$ and different electric fields. The result is depicted in figure (\ref{fig2-MScII}).

The dash lines are the solutions for density of Cooper pairs $\rho/\rho_0=f_3$ as a function of $z/\zeta_{GL}$. They show that even in the presence of an applied magnetic field the increasing of the applied electric field decreases the GL coherence length. The solid lines show the magnetic field in the interior of the superconductor when different electric fields are applied. The curves show that London penetration for the magnetic field decreases. The main conclusion is that under application of an electric field at very low temperatures where there are no normal quasiparticles the critical magnetic field is increased. This can be experimentally tested.

The aim of the present paper was to present the results obtained solving the system of equations, which describes the electrodynamics of s-wave superconductors, for time independent fields and half-plane superconductor geometry. The objective was to elucidate the interplay between applied magnetic and electric fields and superconductivity. The overall conclusion is that the applied magnetic field destroys the superconductivity while the applied electric field supports it. The figures show that when the electric field is applied the density of Cooper pairs $\rho(z)$ increases. Therefore, if we apply electric field at low temperature $\rho(z)$ increases and we can increase the temperature to decrease $\rho(z)$ to the initial value but at higher temperature. Hence, by means of applied electric field we can increase the temperature without destroying the superconductivity.
The results raise new questions. For example the result that at low temperature the applied electric field increases the critical magnetic field makes important the question for the impact of the electric field on Abrikosov vortexes \cite{Abrikosov57,Tinkham75}.

It is important to underline that the scalar and vector fields $Q$ and $\textbf{Q}$ are gauge invariant. This means that they are measurable as the electric and magnetic fields are. The role of these fields is fundamental in superconductivity but not investigated.


\vskip -0.6cm
\begin{thebibliography}{99}
%
\bibitem{London35} F. London and H. London, Proc. R. Soc. London, {\bf Ser. A 149}, 71 (1935).
\bibitem{Hirsch03} J. E. Hirsch, Phys. Rev. {\bf B 68}, 184502 (2003).
\bibitem{Hirsch89} J. E. Hirsch and F. Marsiglio, Phys. Rev. {\bf B 39}, 11515 (1989).
\bibitem{London36} H. London, Proc. R. Soc. London, {\bf Ser. A 155}, 102 (1936).
\bibitem{GL50} V. L. Ginzburg and L. D. Landau, Zh. Eksp. Teor. Fiz. {\bf 20}, 1064 (1950). \
\bibitem{Tinkham75} Michael Tinkham, \emph{Introduction to Superconductivity} (McGRAW-HIL, INC, 1975).
\bibitem{FetterWalecka} Alexander L. Fetter and John Dirk Walecka, \emph{Quantum Theory of Many-Particle Systems}
(McGraw-Hill Book Company,2003).
\bibitem{Hirsch04} J. E. Hirsch, Phys. Rev. {\bf B 69}, 214515 (2004).
\bibitem{Karchev16} N. Karchev, Electrodynamics of s-wave superconductors, arXiv:1512.04284 (2015).
\bibitem{Abrikosov57}A. A. Abrikosov, Zh. Eksperim. i Teor. Fiz., {\bf 32}, 1442 (1957)[Soviet Phys.-JETP, {\bf 5}, 1174 (1957)].


\end{thebibliography}
\end{document}